# Deficiency of the scaling collapse as an indicator of a superconductor-insulator quantum phase transition

Andrey Rogachev[1] and Benjamin Sacépé[2]

[1]Department of Physics and Astronomy, University of Utah, Salt Lake City, UT, USA.,
[2]University Grenoble Alpes, CNRS, Grenoble INP, Institute Néel, 38000 Grenoble, France.

**Finite-size scaling analysis is a well-accepted method for identification and characterization of quantum phase transitions (QPTs) in superconducting, magnetic and insulating systems. We formally apply this analysis in the form suitable for QPTs in 2-dimensional superconducting films to magnetic-field driven superconductor-metal transition in 1-dimensional MoGe nanowires. Despite being obviously inapplicable to nanowires, the 2d scaling equation leads to a high-quality scaling collapse of the nanowire resistance in the temperature and resistance ranges comparable or better to what is accepted in the analysis of the films. Our results suggest that the appearance and the quality of the scaling collapse by itself is not a definitive indicator of a QPT. We have also observed a sign-change of the zero-bias anomaly (ZBA) in the non-linear resistance, occurring exactly at the critical field of the accidental QPT. This behavior is often taken as an additional confirmation of the transition. We argue that in nanowires, the non-linearity is caused by electron heating and has no relation to the critical fluctuations. Our observation suggests that similar to the scaling collapse, the sign-change of ZBA can be a misleading indicator of QPT.**

Quantum phase transitions (QPT) occur at zero temperature between distinct ground states of matter; they are driven by a non-thermal parameter, $g$, which can be, for example, pressure or magnetic field. QPTs take place in many systems ranging from magnetic materials[1,2,3] and superconductors[4,5,6,7] to cold atoms,[8] atomic nuclei[9,10] and stars.[11] The transition from one ground state to another can be of the first order, as in the case of clean metallic ferromagnets.[12] It can also proceed by a smooth evolution of one-ground state to another over broad range of the driving parameter, as in the case of the crossover from Bardeen-Cooper-Schrieffer superconductivity to Bose-Einstein condensation.[13] But, perhaps the most interesting is the case of continuous QPTs, which is characterized by the change of the ground state at certain critical value $g_c$ and by divergence of the correlation length $\xi \propto |g-g_c|^{-\nu}$. Here, $\xi$ has the meaning of the average size of quantum fluctuations and $\nu$ is the correlation length critical exponent.

The dynamics of quantum fluctuations near a QPT are characterized by an additional, temporal length, $\xi_\tau$, related to $\xi$ as $\xi_\tau \propto \xi^z$, where $z$ is the dynamical scaling exponent. At finite temperature, $T > 0$, the maximum possible value for $\xi_\tau$ is $L_\tau \equiv \hbar/kT$ and so the temporal dimension has *finite* length. One can further associate with $L_\tau$ a real space dephasing length, $L_\varphi$, which has a temperature dependence $L_\varphi \propto (\hbar/kT)^{1/z}$. The relation between two length scales $L_\varphi < \xi$ defines the boundary of the critical regime, where the two parameters, $L_\varphi$ and $\xi$, are mutually independent. The theory accounting for finite size in temporal direction is called finite-size scaling.[14] It predicts that in the critical regime, many physical quantities, both thermodynamic and kinetic, assume the scaling form, which for conductivity is expected to follow the equation[14,15,16]

$$\sigma(\delta, T, E) = \frac{Q^2}{\hbar}\left(\frac{k_B T}{\hbar c}\right)^{(d-2)/z} \Phi_\sigma\left(\frac{g-g_c}{T^{1/z\nu}}, \frac{g-g_c}{E^{1/\nu(z+1)}}\right), \qquad (1)$$

Here, $d$ is dimensionality of the system and $Q$ the charge quantum of the order parameter. For a superconductor – insulator transiton, $Q = 2e$; for magnetic systems a characteristic "magnetic charge" can be defined.[16] Finally, $c$ is a non-universal parameter in the prefactor and $\Phi_\sigma$ is non-universal scaling function. The second argument of the scaling function gives the dependence on electrical field, $E$.

Finite-size scaling employing different forms of Eq.1 has been routinely used to analyze the critical regime of the metal – insulator transition,[17,18] the superconductor – insulator transition (SIT)[19,20,21] as well as QPTs in magnetic systems.[22,23,24] More recently, it has been employed to study ultracold atoms in an optical lattice.[25] In this method, one tries to rescale and bring into coincidence the data taken at different $T$ and $g$. The "scaling collapse" and its quality are often taken as an evidence for QPTs. The critical exponents are typically adjustable parameters in this procedure and from their values a universality class can be determined.

For the vast majority of systems tested in experiments, the scaling analysis is applied as a purely phenomenological approach. A question therefore arises: how much can we rely on this method. To address this question, it is instructive to look at systems which allow for a detailed theoretical description of the critical regime and for which the microscopic processes controlling QPTs are well understood. One can then test conclusions drawn from the phenomenological scaling against microscopical critical theory.

There are only few simple *experimentally testable* systems for which such analysis is possible. One notable example is materials with weakly isolated 1d spin chains[26,27] and ladders[28,29] where an exact description is provided by Tomonaga-Luttinger liquid theory and the density matrix renormalization group method.

Recently, we have found that another 1d system, namely superconducting nanowires, undergoes a QPT when driven by a magnetic field, $B$.[30] We have also found that the transition can be completely described by the pair-breaking critical theory with only one adjustable parameter, $C$, in the scaling formula [31]

$$\sigma(B,T) = \frac{(2e)^2}{\hbar}\left(\frac{k_B T}{\hbar D}\right)^{(d-2)/z} \Phi_{1\sigma}\left(C \times F \times \frac{|B^2 - B_c^2|}{B_c^2\ T^{1/z\nu}}\right) . \qquad (2)$$

In this equation, $D$ is the diffusion coefficient, $B_c$ critical field, and $F$ a non-universal constant computed from the parameters of nanowires such as cross-sectional area and critical temperature.[30] The scaling function $\Phi_{1\sigma}$ *is known and numerically computed in the theory*[32] and available for comparison with experiment. The square dependence of the field comes from the pair-breaker dependence $\alpha = kB^2$, combining both orbital and spin terms.

An important distinction between our work on nanowires and previous work on QPTs in superconducting films is that we have used a simple two-fluid model and found that critical behavior occurs only in the superconducting channel. The conductivity of this channel has been approximated as $\sigma_{sc} = \sigma_{ex} - \sigma_{hf}$, where $\sigma_{ex}$ is total experimental conductivity and $\sigma_{hf}$ is the conductivity of normal electrons taken at high fields where superconductivity is suppressed.

Chronologically, before turning to the pair-breaking theory, we applied the phenomenological finite-size scaling analysis to our data following the standard procedure accepted for superconductor – insulator transition in films (thus assuming $d = 2$). Surprisingly, we observed an excellent "data collapse" despite the conspicuous dimensionality inconsistency. We also observed that the accidental transition is accompanied by the sign-change of the non-linear resistance seemingly giving a strong additional argument in favor of QPT. While we now understand that this does not signal any critical behavior, we believe that this misleading detection of a "transition" is instructive and important because it demonstrates that one cannot unambiguously rely on these two indicators of QPT. We present this analysis in the present paper.

In Fig. 1a, we show the resistance versus temperature dependence for nanowire E (we use the same notations as in Ref. [30]) in a parallel magnetic field. The nanowire is made from amorphous Mo$_{78}$Ge$_{22}$ alloy; it has length $L = 3$ μm, width $w = 13$ nm, thickness $t = 6$ nm, and $T_c = 1.5$ K. As Fig. 1a shows, with increasing magnetic field the mean-field critical temperature of the nanowire decreases and then $R(T)$ curves evolve from superconductor-like to insulator-like variation. A more detailed picture of the transitional regime (Fig. 1b) reveals the re-entrant variation at several intermediate fields, namely the resistance first decreases and then increases with temperature, displaying a minimum. A similar re-entrant behavior was also detected in the transverse (perpendicular) field, and in nanowire D, however, for this nanowire the behavior is less pronounced, as shown in Fig. 1c. (Parameters of nanowire D:

composition of alloy $Mo_{50}Ge_{50}$, $L = 3$ μm, $w = 25$ nm, $t = 10$ nm, $T_c = 0.6$ K). We stress that within an interpretation based on the pair-breaking theory, the re-entrant behavior does not carry any special meaning. It appears simply because the resistance of the normal channel, which decreases with temperature, does not completely compensate the increasing resistance of the superconducting channel. (See Supplementary Information to Ref. 30 for more details)

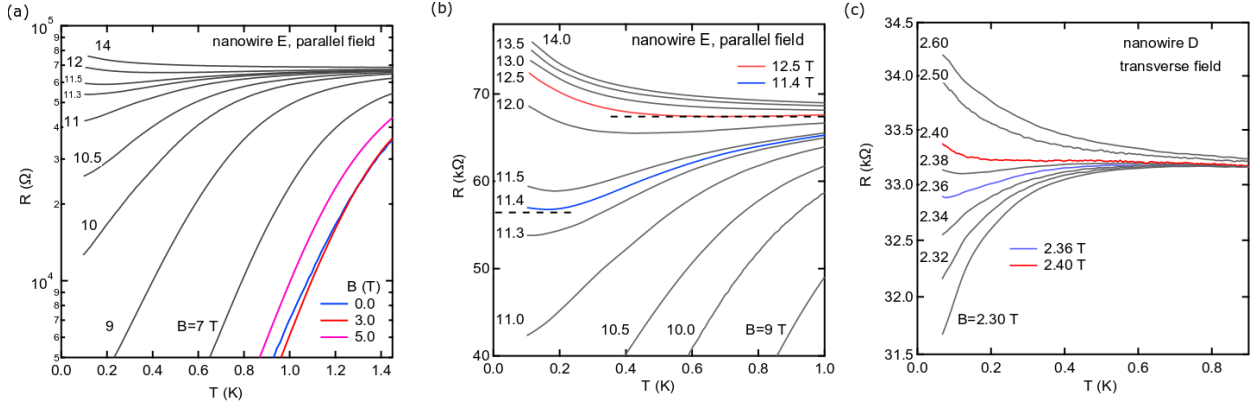

**Fig. 1**. (a) Resistance versus temperature dependence for nanowire E in parallel magnetic field. (b) Same data in the transition range at high fields. The dashed lines indicate resistance at which $R(T)$ changes from superconductor-like to insulator-like behavior at low ($T < 0.25$ K) and higher ($0.4 < T < 1$ K) temperatures. (c) Resistance versus temperature in the transition range for nanowire D in transverse magnetic field.

The peculiar feature of the magnetic-field-induced evolution of the $R(T)$ curves in MoGe nanowires is that it very much resembles the picture of superconductor – insulator transition in thin *films*.[4] For films, the finite-size scaling analysis is most commonly used to scale the resistance. The pre-factor in scaling Eq. 1 is a constant and for low-bias the equation takes the form

$$R(B,T) = R_c \Phi_R \left( \frac{|B - B_c|}{T^{1/z\nu}} \right). \tag{3}$$

The theory supporting this description is the dirty boson model[16] which states that at the transition, the resistance per square is equal to the quantum resistance for pairs, $h/2e^2 \approx 6.5$ kΩ, in certain circumstances of self-duality, but the theory does not provide the scaling function $\Phi_R$ itself. The model implies that at the critical field, the resistance is temperature-independent. Experimentally, a temperature-dependent separatrix between the superconducting and insulating regimes and deviations from 6.5 kΩ have been frequently observed.[4,19,26,21]

We now wish to carry out a formal scaling analysis of our data using Eq.3. Because of the re-entrant behavior, we can identify two temperature-independent separatrices, shown in Fig.1b with dashed lines. The first one appears at lowest temperatures roughly at a field of 11.4 T; the second is in the range $0.4 < T < 1$ K and at a field of 12.5 T. In fact, here we adopt the approach recently used to analyze the magnetic-field-driven SIT in $LaAlO_3/LaTiO_3$ interface superconductors,[32] underdoped $La_{2-x}Sr_xCuO_4$ cuprate,[33] and amorphous WSi films.[34] In all these cases, the observed re-entrant behavior, accompanied by the scaling collapse around each field, was interpreted as a signature of two successive quantum phase transitions.

A standard step in the finite-size scaling analysis of the SIT is to plot the dependence of resistance on the magnetic field (or another driving parameter) at several fixed temperatures. The crossing of the $R(B)$ curves then gives the critical field, $B_c$. In Fig. 2a,b, we plot these dependences measured for nanowire E in parallel fields. As expected, we observe two crossing points, the first at low temperature range $B_{cL} \approx 11.7$ T (Fig. 2a) and the second at high temperature range $B_{cH} \approx 12.5$ T (Fig. 2b). (Similar double-crossing was observed in Refs. [32-34]). The discrepancy between the values of $B_{cL}$ obtained from $R(T)$ and $R(B)$ probably relates to the higher bias current used in the latter measurements, the drift of nanowire resistance, and the time-delay introduced by the inductance of the superconducting magnet.

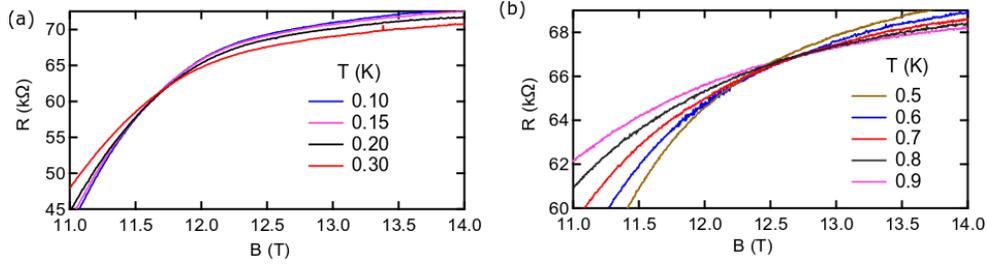

**Fig. 2**. Resistance versus parallel magnetic field for nanowire E at several indicated temperatures. The crossing of the curves indicates two critical fields at lower (a) and higher (b) temperature ranges.

In the next step of the analysis, we plotted the nanowire resistance measured at several fixed temperatures versus scaling variable $|B-B_c|/T^{1/zv}$, as shown in Fig. 3, and used the exponent $zv$ as a parameter in a numerical procedure to minimize the spread of the curves. For both crossing fields, the data show an excellent scaling. The optimized values of the exponents are $zv=4.5$ for $B_{cL}$ (panel a) and $zv=1.15$ for $B_{cH}$ (panel d). The quality of the scaling collapse is in fact much better than what was observed for the majority of the studied superconducting films. Panel (b) of the figure shows the scaled curves near $B_{cL}$ on log-log scale; one can see that the scaling collapse extends along 2 decades along the both axis.

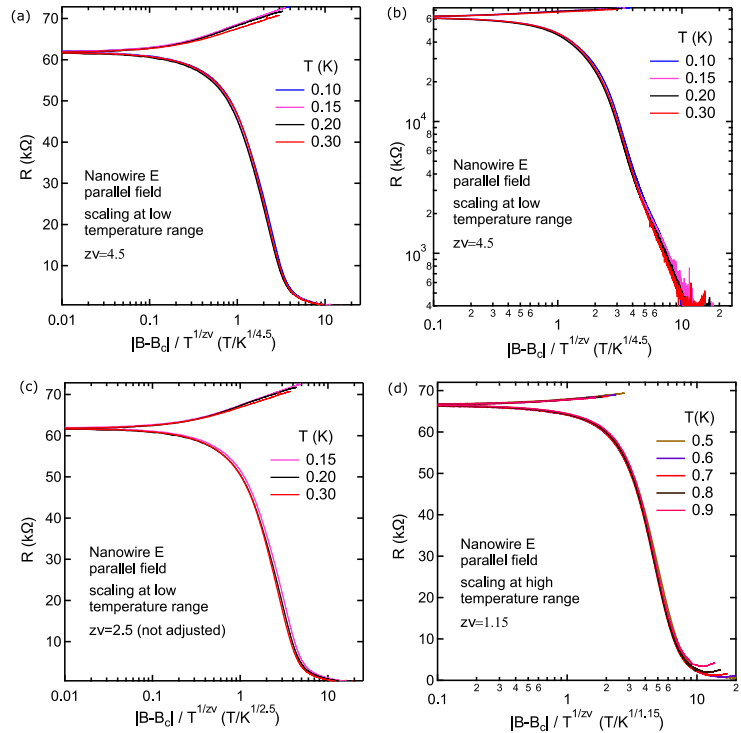

**Fig. 3**. The scaling plot of the resistance of nanowire E in parallel magntic fields. (a) The scaling at low tempertures near the crossing field $B_{cL}=11.7$ T. The exponent has been adjusted to give the best data collapse. (b) The same scaling on log-log scale. (c) The same scaling at with fixed exponent $zv=2.5$ and $R(B)$ curve at T=100 mK not included. (b) The scaling at higher temperatures near the second crossing field $B_{cH}=12.5$ T.

The small upturn seen for the low resistance at higher temperatures (Fig. 3d) appears because of a negative magnetoresistance (MR) contribution. One can indeed notice in Fig. 1a that at lowest temperatures the resistance at field 3 T is actually smaller than the resistance at zero field. Similar negative MR was observed previously in MoGe nanowires fabricated using the molecular template technique.[35] It was attributed to the formation of magnetic atoms on the surface of a wire as a result of its oxidation at ambient conditions. More recently, however, negative MR was detected in LaAlO$_3$/SrTiO$_3$ heterostructures and in amorphous Pb films,[36] where it was observed that the doping of Pb films with magnetic impurities (Cr) actually diminished the magnitude of negative MR. Similarly, we did not detect negative MR in MoGe films doped with Gd.[37] In fact, the effect of magnetic impurities on MoGe appears to be weak; a very large percentage (7 at. %) of Mo atoms were needed to be replaced by Gd to suppress superconductivity, moreover the pair-breaking strength dropped three-fold in thin films.[38] All these recent observations suggest that the interpretation of negative MR as an effect caused by magnetic impurities needs to be revisited.

Similar to nanowire E, we carried out the finite-size scaling analysis for nanowire D in a transverse magnetic field. The scaling plot is shown in Fig. 4a for the low-temperature crossing and in Fig. 4b for the high-temperature crossing. In both cases, the data exhibit the scaling behavior albeit not as perfect as in nanowire E. The scaling collapse appears in a narrow temperature range. Particularly, for high temperature crossing, data at $T = 0.20$ K and at $T \geq 0.50$ K deviate from the "collapsed" curves. The optimized exponents are $zv = 2.5$ for the low-temperature crossing field $B_{cL} = 2.37$ T and $zv = 0.85$ for the high-temperature crossing field $B_{cH} = 2.43$ T.

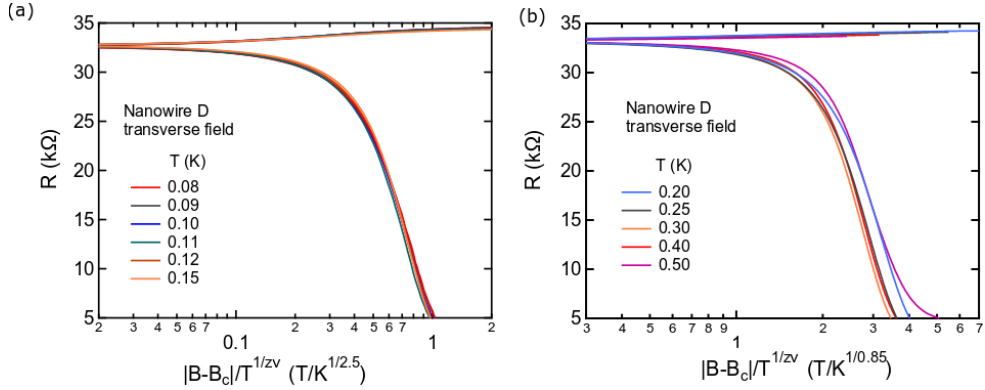

**Fig. 4**. The scaling plot of the resistance of nanowire D in transversel magntic fields. (a) The scaling at low tempertures near the crossing field $B_{cL} = 2.37$ T. (b) The scaling at higher temperatures near the second crossing field $B_{cH} = 2.45$ T.

One can notice that the two exponents for low-field crossing ($zv = 4.5$, for nanowire E and $zv = 2.5$ for nanowire D) are quite different. This could be taken as an evidence for the inconsistency of the whole analysis. However, we found that if we ignore for nanowire E, $R(B)$ curve at lowest temperature, $T = 100$ mK, and use exponent $zv=2.5$, the rest of the data shows fairly good collapse (Fig. 3c). Apparently, even an analysis involving the cross-checking of several samples might still have some ambiguity.

It is instructive at this point to compare the scaling plots obtained under the formal assumption that the dimension of the system is $d = 2$ (Fig. 3,4 above) to those obtained using the proper $d = 1$ scaling equation and the pair-braking critical theory (Fig. 3 in Ref. [30]). The $d = 2$ scaling plots certainly look superior. In the case of nanowires, the $d = 2$ scaling can be rejected because of the dimensional inconsistency. However, in other systems without such an obvious warning, the collapse of data in the finite-size scaling analysis is widely taken for granted to be an indicator of the presence of a QPT. Our on-purpose dimensionality-inconsistent scaling analysis provides a clear demonstration that a scaling analysis, whatever its degree of success, can simply be meaningless.

In this regard the data analysis performed in Ref. [30] within the framework of a pair-breaking QPT substantiates a scaling analysis not simply on the basis of a subjective assessment of a data collapse but with several important additional facts: (i) No adjustment of the exponents was done; the exponents predicted by the theory, $z \approx 2$ and $\nu \approx 1$, were used in comparison with experiment. (ii) The predicted value of $\Phi_\sigma(0) = 0.218$ is reproduced within a factor of two. (iii) For different wires and field orientations, the conductivity followed the theoretically predicted scaling function of the pair-breaking QPT theory with the same adjustable parameter $C$. (iv) The pair-breaking QPT occurs only in the superconducting part of the system and conductance due to critical superconducting fluctuations is about 10% of that of normal electrons at $B_c$. The critical contribution was singled out using an approximate two-fluid model, which explains the deviations from perfect scaling.

The case of superconducting nanowires (and several other 1-d systems) is special since there is an exact description of the critical regime. The theoretical description of QPTs in $d = 2,3$ dimensions is much more challenging. Still, the example of superconducting nanowires shows that it is highly desirable to have at least some verifiable theoretical predictions beyond the finite-size scaling.

Let us now discuss the variation of the non-linear resistance across the transitional regime. In Fig. 5a, we plot the dependence of the differential resistance versus current at fixed transverse magnetic fields for nanowire D. The differential resistance displays a zero-bias anomaly which changes sign from negative to positive at a magnetic field $B \approx 2.38$ T. This value coincides very well to the field at which the $R(T)$ variation becomes temperature-independent (Fig.1c). Similar behavior was detected for nanowire E in a transverse magnetic field. For this nanowire, the change from a superconductor-like to an insulator-like variation of $R(T)$ curves at $B \approx 5.4$ T (Fig. 5b) agrees well with the field $B \approx 5.35 - 5.4$ T at which the zero-bias anomaly changes sign from negative to positive (Fig. 5c).

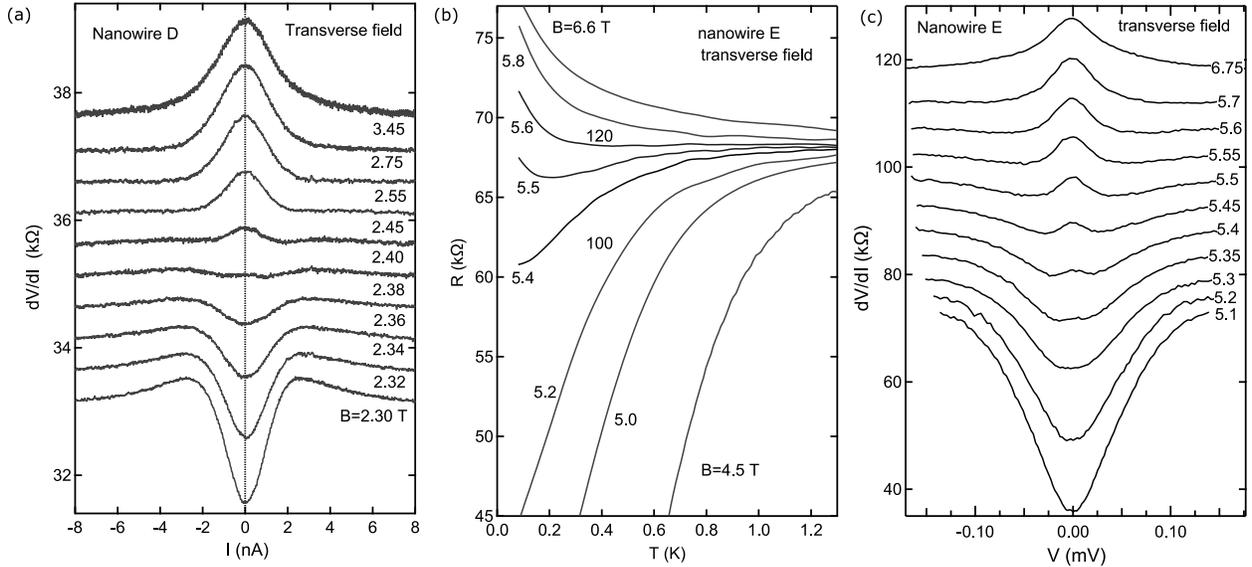

**Fig. 5**. (a) Differential resistance versus current at indicated magnetic transverse fields for nanowire D at $T = 80$ mK. For clarity, the curves are upshifted in 500 Ω increment. The zero bias anomaly changes sign from negative to positive at $B \approx 2.38$ T. (b) Resistance versus temperature for nanowire E in a transverse magnetic field. (c) Differential resistance versus voltage for nanowire E at indicated magnetic fields and at $T = 80$ mK. For clarity, the curves are upshifted in 5 kΩ increments.

In the past, such concurrent variations of linear and non-linear resistance have been taken as evidence for superconductor-insulator transitions in resistively-shunted single Josephson junction,[39] one-dimensional (1d) Josephson junction arrays,[40] superconducting nanowires[41,42] and films,[43,44] and for the metal-insulator transition in multilayer $MoS_2$.[45] This variation is indeed expected from the general scaling equation, Eq.1, which states that at the critical field and fixed temperature the resistance of a system should not depend on electrical field, regardless of the system dimension.

In our case of the magnetic-field-driven QPTs in superconducting nanowires, the situation is different. The pair-breaking critical theory affirms that the critical behavior occurs in the superconducting channel only. Moreover, at the critical field the contribution of the superconducting fluctuations is temperature-dependent, $\sigma_c \propto T^{-1/z}$. We therefore expect that there must be some non-critical processes which make the change of sign of the ZBA coincidental with the temperature-independent separatrix in the $R(T)$ curves.

The negative ZBA at low fields (in the superconducting regime) naturally occurs because the increasing current reduces the density of Cooper pairs. We recently showed that at high magnetic fields, where superconductivity is completely suppressed, the positive ZBA occurs due to electron heating.[46] The heating model semi-quantitatively explained both the height and width of the anomaly as well as its dependence on nanowire length. Moreover, in Ref. [46] we fabricated and studied several special nanostructures, which allowed us to reject Coulomb blockade, the Altshuler-Aronov anomaly, and non-uniform nanowire morphology as possible sources of the positive ZBA.

Electron heating, combined with the two-fluid model, gives a natural explanation of the concurrent behavior of the linear and non-linear resistance in MoGe nanowires. The temperature-independent segment of $R(T)$ occurs because at $B_{cL}$, the decreasing (with temperature) resistance in the normal channel is completely compensated for by increasing resistance in superconducting channel. In the first approximation, the effect of the electrical field on both channels is equivalent to raising temperature and hence *E*-independent behavior is expected in the $dV/dI$ curve. Shifting the magnetic field away from $B_{cL}$ breaks the compensation between the conductance of the normal and superconducting channels and produces the insulator-like or superconductor-like $R(T)$ variation accompanied by the positive or negative ZBA.

Let us summarize our main observation and conclusion. We have formally employed the phenomenological scaling equation commonly used in the finite-size scaling analysis of superconducting films ($d=2$) to superconducting nanowires ($d=1$). Despite being obviously inapplicable, the minimization procedure returns a very good scaling collapse of the data, which provides reasonable values of the critical exponents and covers a range of temperatures and resistances compatible with what is accepted for the analysis of SIT in films. The scaling behavior is accidental and misleading, yet it has been observed in our all measurements. We certainly do not dispute the theoretical picture that near QPT scaling behavior is present; neither we claim that in all works cited in this article the detection and characterization of QPT is incorrect. Still our work demonstrates that scaling analysis has a deficiency and observation of the scaling collapse by itself does not constitute a definitive proof for the occurrence of a QPT. It is highly desirable to go beyond finite-size scaling and have theoretical models that provide additional verifiable parameters such as for example values of critical exponents and parameters that would allow cross-checking between different samples and making connection to known or experimentally-measurable quantities of a system under study.

**Acknowledgments**
This work has been supported by NSF grants DMR1611421 and DMR1904221. B.S. has received funding from the European Research Council (ERC) under the H2020 program (grant No. 637815) and from the French National Research Agency (ANR grant *CP-Insulator*).